# Optical super-resonances in dielectric microsphere particles

Zengbo Wang[*a], Boris Luk'yanchuk[b], Baidong Wu[a], Bing Yan[a], Ahmetova Assel[b], Igor Yaminsky[b], Haibo Yu[c], Lianqing Liu[c]

[a]School of Computer Science and Electronic Engineering, Bangor University, Dean Street, Bangor, Gwynedd, LL57 1UT, UK; [b]Faculty of Physics, Lomonosov Moscow State University, Moscow 119991, Russia; [c]State Key Laboratory of Robotics, Shenyang Institute of Automation, Chinese Academy of Sciences, Shenyang 110016, China

## ABSTRACT

Extreme field localization and giant field enhancement are often achieved by using plasmonic nanostructures and metamaterials such as strongly coupled silver nanoparticles. Dielectric particles and structures can focus light beyond the diffraction limit (photonic nanojet effect), but with much weaker strengths. Recently, we showed that dielectric microspheres could support high-order Mie resonance modes ('super-resonance modes'), that can generate similar level of electric field intensity enhancement as plasmonic structures on the order of $10^4$- $10^7$.

In this work, we aim to further advance our understanding of the super-resonance modes. New results on the effects of size parameter and refractive index on optical super-resonances across a wide parameter range and with improved numerical accuracies is presented. The results suggest that the electric field intensity enhancement could reach a record high level of $10^9$-$10^{11}$ at specific conditions that surpass plasmonic enhancements. Moreover, super-resonance-enabled focusing by microsphere lens under different lighting sources (e.g., different color LEDs or lasers, halogen lamps) is investigated and compared for the first time. These results are important in understanding the super-resolution mechanism for microsphere nanoscopy and will find numerous potential applications in photonics.

## 1. INTRODUCTION

When light interacts with transparent microparticles, subwavelength focusing and field localization could be achieved at the shadow-side of the microparticles, including micro and nano spheres, prisms, cubes and even biological cells and spider silks. This effect is known as 'Photonic Nanojet' (PNJ), which was first discovered in 2000 when researchers worked on laser cleaning project and found a subwavelength hole often appear on substrate after laser removal of microspheres[1]. The effect was consequently used for large-area laser surface nanopatterning and rapidly expanded into a variety of other areas including super-resolution imaging, microscopy, fabrication, sensing, trapping, sorting, manipulation, and signal amplification (e.g., Raman, photoluminescence), and extended to the acoustic and terahertz domain [2-8].

In most cases, the lateral resolution of PNJ focusing is close to or slightly better than the diffraction limit in air ($\lambda/2$), which is insufficient to explain the experimental super-resolution of $\sim \lambda/7$ obtained in microsphere-assisted super-resolution imaging (aka. Microsphere nanoscopy) [9-15]. Besides, in laser cleaning/nanopatterning experiments, researchers can occasionally see strong explosion of some microspheres which broken into tiny pieces. These observations suggest there could be another field confinement mechanism takes place beyond the PNJ. In 2019, we reported that weakly dissipating dielectric spheres made of materials such as glass, quartz, etc. can support 'super-resonance modes' associated with internal Mie modes. These resonances, happening for specific values of the size parameter, yield field-intensity enhancement factors on the order of $10^4$–$10^7$, which can be directly obtained from analytical Mie calculations [16]. This provides a new mechanism to obtain extreme field localization with giant field enhancement that is comparable to reported field enhancement by plasmonic nanostructures and metamaterials such as strongly coupled silver nanoparticles. It also provides a new theoretical route to explain the deep super-resolution observed in microsphere nanoscopy technique [9].

*z.wang@bangor.ac.uk; phone +44(0) 1248 382696; fax +44(0) 1248 362686

Super-resonances modes are extremely sensitive to the size parameter *q* of microspheres, defined as q=2πa/λ, where a is the radius of particle and λ the incident wavelength. We have been using a *q* sampling accuracy of *dq=1e-4* in previous studies [9, 16-18]. This has enabled us to find super-resonance modes with giant field enhancement up to $10^7$. However, as shown in this work, the 1e-4 sampling accuracy could become insufficient for large q values and a much higher sampling accuracy (up to 1e-14) is needed to precisely find and simulate the super-resonance modes. With improved sampling accuracy, we observed new super-resonance modes that are not seen in previous calculations, with record high field enhancement of the level of $10^9$-$10^{11}$. Besides, we also show that these super-resonance modes have important effects on microsphere super-resolution imaging when different-spectrum light sources were used.

## 2. METHODS

Mie theory is used, which express the incident, scattering and internal fields in the forms of superpositions of partial waves in terms of spherical harmonics. Details of Mie formulation can be found in [19]. Because super-resonance modes are excited internal partial wave modes, it is useful to introduce a new efficiency term 'internal scattering efficiency, $Q_{in}$' not defined in standard Mie theory:

$$Q_{in} = \sum_{\ell=1}^{\infty}\left(Q_\ell^{(e)} + Q_\ell^{(m)}\right), \quad Q_\ell^{(e)} = \frac{2}{q^2}(2\ell+1)|c_\ell|^2, \quad Q_\ell^{(m)} = \frac{2}{q_m^2}(2\ell+1)|d_\ell|^2, \quad (1)$$

where

$$c_\ell = \frac{in_p}{\Re_\ell^{(a)} + i\Im_\ell^{(a)}}, \quad d_\ell = \frac{in_p}{\Re_\ell^{(b)} + i\Im_\ell^{(b)}}. \quad (2)$$

are internal wave amplitudes, with $\Re_\ell^{(a,b)}$ and $\Im_\ell^{(a,b)}$ coefficients expressed by

$$\begin{aligned}
\Re_\ell^{(a)} &= n_p\psi_\ell(q_p)\psi_\ell'(q_m) - n_m\psi_\ell(q_m)\psi_\ell'(q_p), \\
\Im_\ell^{(a)} &= n_p\psi_\ell(q_p)\chi_\ell'(q_m) - n_m\chi_\ell(q_m)\psi_\ell'(q_p), \\
\Re_\ell^{(b)} &= n_p\psi_\ell(q_m)\psi_\ell'(q_p) - n_m\psi_\ell(q_p)\psi_\ell'(q_m), \\
\Im_\ell^{(b)} &= n_p\chi_\ell(q_m)\psi_\ell'(q_p) - n_m\chi_\ell'(q_m)\psi_\ell(q_p).
\end{aligned} \quad (3)$$

Here the functions $\psi_\ell(z) = \sqrt{\frac{\pi z}{2}} J_{\ell+\frac{1}{2}}(z)$ and $\chi_\ell(z) = \sqrt{\frac{\pi z}{2}} N_{\ell+\frac{1}{2}}(z)$ are expressed through the Bessel and Neumann functions.

First, a Fortran-based Mie program was written to compute the $Q_{in}$ efficiency as a function of *q* within range 0<q<100. A sampling accuracy of *dq=1e-4* was first used. The calculated $Q_{in}$ data was fed through a peak-finding algorithm which finds resonance peaks. Then, a local-peak-refining program was developed: for each resonance peak, the refining program will repeat $Q_{in}$ calculation around that peak location with improved *dq/10* sampling accuracy; such refining process is repeated until the final step accuracy *dq*=1e-14 is reached. As a result, we obtained super-resonance modes plot calculated at different local sampling accuracies, from 1e-4, 1e-5, to 1e-6 and 1e-7 and so on until 1e-14. All these data are plotted and compared which reveals new physics previously unseen.

Once accurate super-resonance peaks are found, we proceed to calculate the electric and magnetic intensity field distribution in XZ plane for each super-resonance modes (incident beam *x*-polarized, propagates from *-z* to *+z* direction). The calculation was performed with spatial resolution a/50 within XZ plane ranging from -1.6a to 1.6a, where a is radius of particle. The maximum value of the field intensity in the 2D plane was then found and compared.

To investigate incident light spectrum effect on microsphere focusing, each incident wavelength (converted to *q*) will be modulated by a weighting function *w(q)* given by the normalized light spectrum. The total electric field intensity is then computed as:

$$I(q;x,z) = \frac{\int_{q_1}^{q_2}|E|^2(q;x,z).w(q)dq}{\int_{q_1}^{q_2}w(q)dq}$$

(4)

Similarly for total magnetic field intensity:

$$I(q;x,z) = \frac{\int_{q_1}^{q_2}|H|^2(q;x,z).w(q)dq}{\int_{q_1}^{q_2}w(q)dq}$$

(5)

The numerical integration was undertaken by trapezoid approximation method using 1e-4 step accuracy, but with all super-resonance modes identified at 1e-14 accuracy inserted during integration. The process leads to a non-constant spacing and we use approximation equation:

$$\int_{q_1}^{q_2} I(q;x,z)dq \approx \frac{1}{2}\sum_{n=1}^{N}(q_{n+1}-q_n)[I(q_n;x,z)+I(q_{n+1};x,z)]$$

(6)

Such computing is highly intensive and performed by parallelized Mie code (10 parallel threads). Typical calculation takes about 3-6 hours, depending on the particle size and light spectrum.

## 3. RESULTS

### 3.1 Effect of sampling accuracy

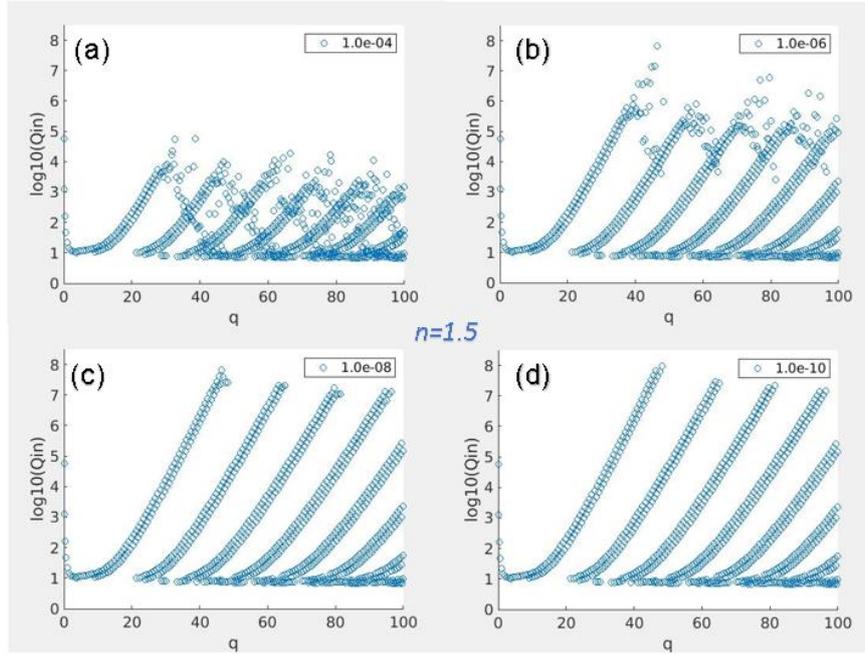

Figure 1. Resonance peaks found at different *q* sampling accuracy *dq* for a n=1.5 microsphere. (a) *dq*=1e-4, (b) *dq*=1e-6, (c) *dq*=1e-8, (d) *dq*=1e-10. Note the y-axis is plotted in logarithmic scale.

Figure 1 shows the evolution of resonance peaks calculated at different *dq* sampling accuracy, As sampling accuracy increases (*dq* decrease), the resonance points become more clearly aligned and the peak value increased from $Q_{in} \approx 10^5$ (Fig.1a) to $Q_{in} \approx 10^8$ (Fig.1d). The high accuracy sampling process is thus essential to find all super-resonance peaks which wasn't demonstrated in previous literature.

## 3.2 Extreme field enhancement

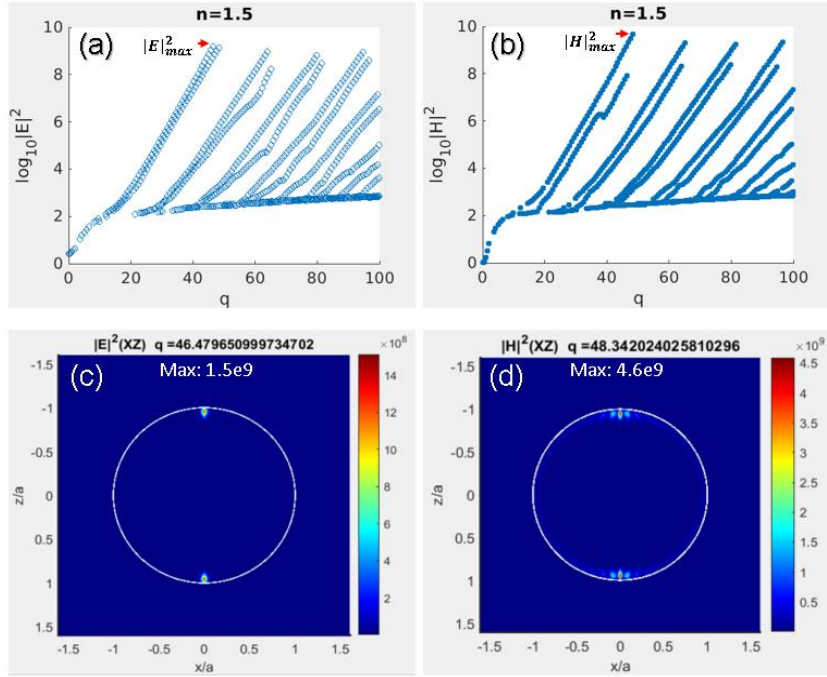

Figure 2. Giant field enhancement for a super-resonant n=1.5 microsphere. (a,c) q=46.47965099973470, $|E|^2$=1.52e9. (b,d) q=48.3420240258103, $|H|^2$=4.60e9

The refining process allows precision location of peak super-resonance modes. The field distribution was then calculated based on peak locations in Fig. 1(d), and corresponding results was shown in Fig.2. The maximum electric field enhancement reaches $|E|^2$=1.52e9 at q=46.47965099973470 (Fig.2a, 2c), and magnetic field $|H|^2$=4.60e9 at q=48.3420240258103 (Fig.2b, 2d). Two almost symmetrical hotspots appear at the top and bottom apex of the microsphere along the light propagation direction, indicating such modes are strongly coupled to a circulation mode within the spherical cavity. A further analysis reveals the excited partial wave are $l = 63$ mode for Fig.2a and 2c, and $l = 65$ mode for Fig.2b,2d. It is also interesting to note the maxim values appear to repeat periodically with a period around $dq \approx 15 - 17$ towards larger q range, but its values are either close to or slightly smaller than the first peak value. Such periodicity may provide a new route for experimental selection of microspheres to achieve best field location and improved focusing resolution.

## 3.3 Refractive index effect

It is important to see how super-resonance modes change with refractive index of microspheres. Table 1 summarizes calculated peak super-resonance modes for n=1.5, 1.9, 2.4 and 4.0, respectively. Two tendencies: peak enhancement increase with refractive index, at n=4.0, peak field enhancement reaches 1.0e11 order. Second, domain mode index, $l$, decreases with refractive index, from order $l$ = 63-65 for n=1.5 to order $l$ =15-18 for n=4.

Table 1. List of super-resonance peaks.

| n | q | $E^2$ | $H^2$ | Mie Mode ($l$) |
|---|---|---|---|---|
| 1.5 | 46.47965099973470 ($E^2$) | 1.52e9 | --- | $d_{63}$ |
|  | 48.3420240258103 ($H^2$) | --- | 4.60e9 | $c_{65}$ |
| 1.9 | 21.840288946319699 | 7.89e9 | 2.60e10 | $c_{35}$ |
| 2.4 | 12.049917983360400 | 1.74e10 | 9.20e10 | $c_{23}$ |
| 4.0 | 8.027975095974320 ($E^2$) | 1.05e11 | --- | $c_{18}$ |
|  | 6.182621859591350 ($H^2$) | --- | 6.32e11 | $c_{15}$ |

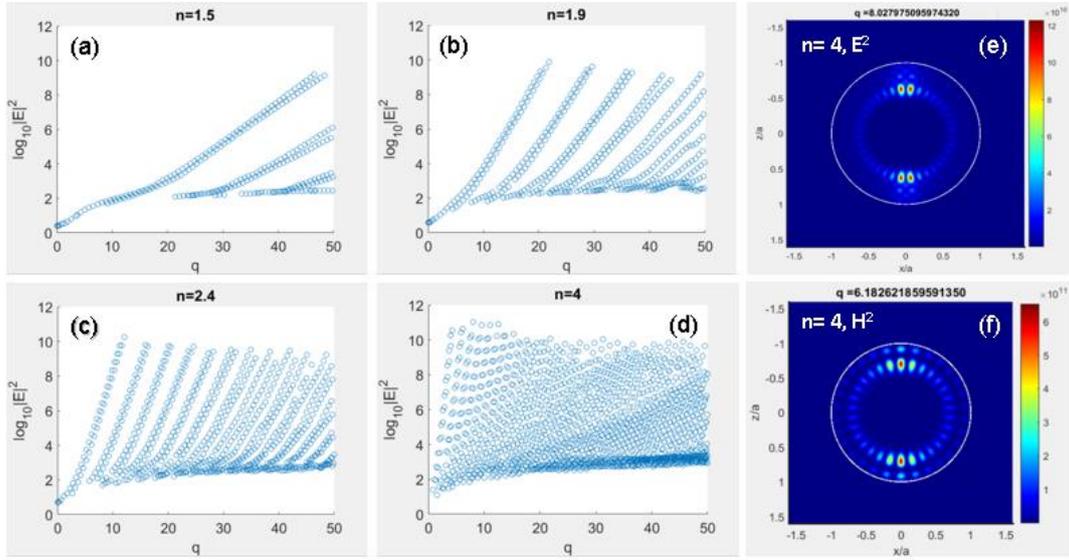

Figure 3. Electric field enhancement versus q for different refractive index particles. (a) n=1.5, (b) n=1.9, (c) n=2.4 and (d) n=4.0. (e) Electric intensity field distribution for n=4, q=8.02797509597432, $E^2$=1.05e11. (f) Magnetic intensity field distribution for n=4, q=6.182621859591350, $H^2$=6.32e11;

From Fig.3, we can see that, as refractive index increases, the slope of the peak curve also increases, leading to increased number of super-resonance modes being found within a given size parameter range. In other words, it might be easier to observe super-resonance modes in higher-index microspheres. However, the localized fields for higher-index particle are located well inside the particle (Fig.3e, 3f), not like n=1.5 particle whose field localization is near to particle surface (Fig.2) thus accessible for wider range of applications.

### 3.4 Broadband light sources

In microsphere nanoscopy imaging, a broadband halogen lamp (e.g., [10, 12]) is often used, other types of light sources including different color LED or lasers have also been used, including for example 405 nm, 532 nm, and 671 nm lasers in [20], and 633 nm laser in [21]. Majority of theoretical treatment takes peak wavelength as simulation wavelength and neglects other wavelengths. Very recently, Mandal et al. studied and considered all wavelengths in different broadband spectrums on microsphere focusing, including halogen lamp, white LED, super-continuum laser, Hg arc lamp. They found small difference exists in investigated cases[22]. However, their study has not taken super-resonance modes into account which are presented within the incident spectrum. Here, we demonstrate how light is focused by n=1.5 particles under different light spectrums but with all super-resolution modes being included, as shown in Fig.4.

From Fig.4, we can see for a n=1.5, d=5 μm, microsphere, the Blue LED source (Fig.4a, $q_{peak} = 39.2699$ at $\lambda = 400\ nm$) generates stronger focusing and field enhancement due to excitation of more pronounced super-resonance modes compared to green LED (Fig.4b, $q_{peak} = 29.9199$ at $\lambda = 525\ nm$)) and red laser (Fig.4c, $q_{peak} = 24.7370$ at $\lambda = 635\ nm$)) light sources. Blue light is thus recommended for improving the imaging resolution when a 5-μm microsphere lens as used. Meanwhile, Fig.4 (d1-d4) show light focusing under halogen lamp but with particle size, from 5 μm to 8 μm. As it can see, as particle size increase, stronger focusing can be generated due to more pronounced super-resonance modes are excited, which leads to improved focus resolution and thus beneficial for super-resolution imaging. The finding is different from conventional single wavelength simulation results in the literature which suggests smaller particle has tinier focusing and higher resolution [10]. Such discrepancy would need more careful experimental research in the future. For experimental studies, the challenge is that these super-resonance modes are highly sensitive to imperfection in microspheres, including geometry and material loss. A larger microsphere may become less possible to achieve the perfect spherical geometry and zero material loss conditions for predicted super-resonance effect here. In next step work, we will consider material loss and dispersion in microsphere materials so that a more realistic design of microsphere and lighting source can be selected for optimized super-resolution imaging applications.

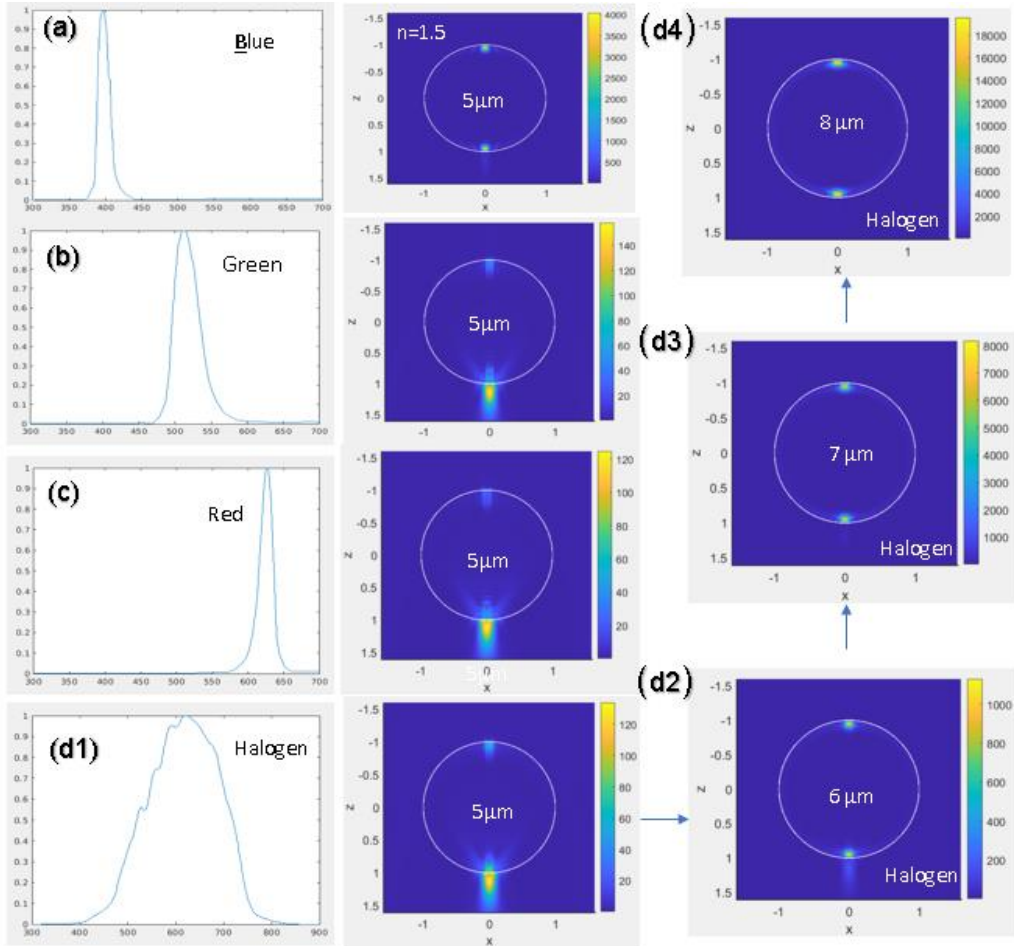

Figure 4. Light focusing by a n=1.5 microsphere illuminated by different spectrum light sources. (a) Blue LED source, (b) Green LED, (c) Red laser, (d1-d4) Halogen lamp but with different particle sizes from 5 μm to 8 μm.

## 4. CONCLUSIONS

In this work, we have investigated super-resonance modes in microspheres by using an improved algorithm with numerical accuracy down to 1e-14. The results suggest that the electric field intensity enhancement could reach a record high level of $10^9$-$10^{11}$ at specific conditions for n=1.5, 1.9, 2.4, 4.0 particles. Moreover, super-resonance-enabled focusing by microsphere lens under different lighting sources (e.g., different color LEDs or lasers, halogen lamps) have been investigated. Blue light source is found to be beneficial for super-resolution imaging. These results are important in understanding the super-resolution mechanism and will provide a new route in optimization microsphere systems for microsphere nanoscopy. Moreover, the giant field enhancement will find potential applications in super-enhanced Raman scattering, non-linear optics and others.

We acknowledge funding support from European ERDF grants (SPARCII c81133), UK Royal society (IEC\R2\202178 and IEC\NSFC\181378) and Russian Foundation for Basic Research (no. 21-58i-10005).